# Oxygen hyperstoichiometric hexagonal ferrite $CaBaFe_4O_{7+\delta}$ ($\delta \approx 0.14$) : coexistence of ferrimagnetism and spin glass behaviour

---

Tapati Sarkar*, V. Duffort, V. Pralong, V. Caignaert and B. Raveau

*Laboratoire CRISMAT, UMR 6508 CNRS ENSICAEN,*

*6 bd Maréchal Juin, 14050 CAEN, France*

**Abstract**

An oxygen hyperstoichiometric ferrite $CaBaFe_4O_{7+\delta}$ ($\delta \approx 0.14$) has been synthesized using "soft" reduction of $CaBaFe_4O_8$. Like the oxygen stoichiometric ferrimagnet $CaBaFe_4O_7$, this oxide also keeps the hexagonal symmetry (space group: $P6_3mc$), and exhibits the same high Curie temperature of 270 K. However, the introduction of extra oxygen into the system weakens the ferrimagnetic interaction significantly at the cost of increased magnetic frustration at low temperature. Moreover, this canonical spin glass ($T_g$ ~ 166 K) exhibits an intriguing cross-over from de Almeida-Thouless type to Gabay-Toulouse type critical line in the field temperature plane above a certain field strength, which can be identified as the anisotropy field. Domain wall pinning is also observed below 110 K. These results are interpreted on the basis of cationic disordering on the iron sites.

* Corresponding author: Tapati Sarkar

e-mail: tapati.sarkar@ensicaen.fr

## Introduction

The recent discovery of the new series of “114” oxides, (the cobaltites – $(Ln,Ca)_1BaCo_4O_7$ [1 – 8] and the ferrites – $(Ln,Ca)_1BaFe_4O_7$ [9 – 11]) have opened up a new field for the investigation of strongly correlated electron systems. These oxides consist of $CoO_4$ (or $FeO_4$) tetrahedra sitting in alternating layers of kagomé and triangular arrays [10]. The structure can also be described as the stacking of close-packed $[BaO_3]_\infty$ and $[O_4]_\infty$ layers whose tetrahedral cavities are occupied by $Co^{2+}/Co^{3+}$ (or $Fe^{2+}/Fe^{3+}$) species, forming triangular and kagomé layers of $CoO_4$ (or $FeO_4$) tetrahedra. This structure has been primarily responsible for the wide variety of magnetic states that has been observed in this group of oxides, ranging from a spin glass for cubic $LnBaFe_4O_7$ [9, 10] to a ferrimagnet for orthorhombic $CaBaCo_4O_7$ [5] and hexagonal $CaBaFe_4O_7$ [9] oxides.

Recent studies of the “114” cobaltites [12 – 17] have revealed the existence of closely related structures with various crystallographic symmetries, and possibility of oxygen non-stoichiometry in the range “$O_7$” to “$O_{8.5}$” in those systems. This change of oxygen stoichiometry, which induces the variation of $Co^{2+}$:$Co^{3+}$ ratio in the system, is expected to influence the physical properties of these compounds considerably. This is the case of the oxygen rich “114” cobaltites $YBaCo_4O_{8.1}$ [15] and $YbBaCo_4O_{7.2}$ [17], which were shown to be magnetically frustrated rather than magnetically ordered at low temperatures.

In contrast to the cobalt oxides, no report of oxygen hyperstoichiometric “114” ferrites exists till date, probably due to the fact that $Fe^{2+}$ gets too easily oxidized into $Fe^{3+}$, thereby destabilizing the “114” structure at the benefit of pure “$Fe^{3+}$” oxides. We have, thus, investigated the possibility to stabilize the mixed valence $Fe^{2+}/Fe^{3+}$ in the “114” oxygen hyperstoichiometric $CaBaFe_4O_{7+\delta}$ ferrite by reducing the fully oxidized compound $CaBaFe_4O_8$ [18] at low temperature in an argon-hydrogen atmosphere. We report herein on the magnetic properties of the “114” oxygen hyperstoichiometric $CaBaFe_4O_{7.14}$ hexagonal ferrite. We show that, like the stoichiometric phase $CaBaFe_4O_7$, this oxide also exhibits ferrimagnetism with a $T_C$ of 270 K, but that the competition between ferrimagnetism and magnetic frustration is much more pronounced than for the stoichiometric phase, as seen from the decrease of the magnetization. More importantly, we observe that $CaBaFe_4O_{7.14}$ is characterized by a canonical spin glass behaviour with $T_g \approx 166$ K, and an intriguing cross-over from an Ising to a Heisenberg spin glass type behaviour in the external magnetic field at low temperature. Besides this competition between ferrimagnetism and spin glass behaviour, one also observes domain wall pinning below 110 K. This very different magnetic behaviour of $CaBaFe_4O_{7.14}$ is explained in terms of cationic deficiency and disordering on the iron sites, the "barium-oxygen" hexagonal close packing remaining untouched.

## Experimental

The precursor $CaBaFe_4O_8$ [18] was prepared by the sol gel method. Stoichiometric amounts of calcium carbonate (Prolabo, 99%) and barium carbonate (Alfa Aesar, 99%) were dissolved in a large excess of melted citric acid monohydrate at ~ 200°C. Iron citrate (Alfa Aesar, 20% of Fe) was separately dissolved in hot water leading to a dark brown solution which was poured on the citrate mixture. The water was then evaporated followed by decomposition of the gel. The gel was calcined at 450 °C under air to obtain an amorphous precursor, which was then pressed into pellets before firing at 1200 °C to obtain $CaBaFe_4O_8$.

The oxygen hyperstoichiometric "114" ferrite, $CaBaFe_4O_{7+\delta}$ was then obtained by reducing $CaBaFe_4O_8$ under an $Ar/H_2$ 10% mix at 610 °C for 24 hrs.

The oxygen content of the sample was determined by redox titration. The sample was dissolved in hot HCl (3M) flushed with argon to remove the dissolved oxygen. After cooling down the solution, $Fe^{2+}$ cations were titrated using $2 \times 10^{-2}$ M cerium(IV) sulfate (Riedel-de Haën) and 1.10-phenantroline iron(II) sulfate (Alfa Aesar) as an indicator under constant argon flow. We obtained $\delta = 0.14$.

The X-ray diffraction patterns were registered with a Panalytical X'Pert Pro diffractometer under a continuous scanning mode in the 2θ range 10° - 120° and step size $\Delta 2\theta$=0.017°. The d.c. magnetization measurements were performed using a superconducting quantum interference device (SQUID) magnetometer with variable temperature cryostat (Quantum Design, San Diego, USA). The a.c. susceptibility, $\chi_{ac}(T)$ was measured with a Physical Property Measurement System (PPMS) from Quantum Design with the frequency ranging from 10 Hz to 10 kHz ($H_{dc}$ = 0 Oe and $H_{ac}$ = 10 Oe). All the magnetic properties were registered on dense ceramic bars of dimensions ~ 4 × 2 × 2 $mm^3$.

## Results and discussion

*Structural Characterization*

The X-ray diffraction pattern (Fig. 1) revealed that $CaBaFe_4O_{7.14}$ stabilized in the same hexagonal symmetry (space group: $P6_3mc$) as the "$O_7$" phase [9]. The Rietveld analysis from the XRD data was done using the FULLPROF refinement program [19]. The fit is also shown in Fig. 1 (red curve). The bottom blue curve corresponds to the difference between the observed and calculated diffraction patterns. Satisfactory matching of the experimental data with the calculated profile of the XRD pattern and the corresponding reliability factors $R_F$ = 3.88 % and $R_{Bragg}$ = 5.01 % confirm that the fit obtained is reasonably accurate. The extracted lattice parameters ($a$ = 6.355 Å, $c$ = 10.372 Å) show a very marginal increase over the "$O_7$" phase – $a$ increases by ~ 0.11 % while $c$ remains virtually unchanged. The refinements of the atomic coordinates, thus, lead to results similar to those previously obtained for $CaBaFe_4O_7$ [9]. The low $\delta$ value and the low scattering factor of oxygen do not allow any oxygen excess or cationic deficiency to be detected from X-ray powder diffraction data. A very careful neutron diffraction study might perhaps allow the issue to be sorted out, but will really be at the limit of accuracy, and consequently, is not within the scope of this paper.

*D. C. magnetization study*

In Fig. 2, we show the Zero Field Cooled (ZFC) and Field Cooled (FC) magnetization of $CaBaFe_4O_{7+\delta}$ recorded under a magnetizing field of 0.3 T. The sample shows the same increase in magnetization below ~ 270 K as the oxygen stoichiometric oxide indicating a similar transition to an ordered magnetic state below 270 K. However, a careful look at the magnetization values reached at the lowermost measured temperature (5 K) immediately reveals a striking difference in the magnetic behaviour of $CaBaFe_4O_{7.14}$ vis – à – vis that of $CaBaFe_4O_7$. While the F.C. magnetization of the oxygen stoichiometric compound reaches a value of more than 2.5 $\mu_B$/f.u. at T = 5 K, the maximum magnetization value of our oxygen rich sample is only 0.93 $\mu_B$/f.u., which is less by more than a factor of ½.

This large difference in the magnetization value at low temperature between the two samples ($\delta = 0$ and $\delta > 0$) prompted us to record the hysteresis curve of our oxygen rich sample at low temperature (T = 5 K) and compare it with that of the oxygen stoichiometric sample. This is shown in Fig. 3. The magnetization value obtained at the highest measuring field of 5 T (2.5 $\mu_B$/f.u.) is again different from the oxygen stoichiometric sample (3.1 $\mu_B$/f.u.), as expected. More importantly, a rather striking difference is seen in the shape of the hysteresis loop. The coercive field ($H_C$) and remanent magnetization ($M_r$) of our sample at T = 5 K are 0.77 T and 0.63 $\mu_B$/f.u. respectively. While the value of the coercive field compares well with that of the oxygen stoichiometric sample, the value of the remanent magnetization is much lower than that obtained for $CaBaFe_4O_7$ ($M_r$ ~ 1.8 $\mu_B$/f.u.). This results in the overall shape of the hysteresis loop of our sample (Fig. 3) to be very different from that of the oxygen stoichiometric sample (inset of Fig. 3). The degree of magnetic saturation in a sample can be roughly quantified from the M-H loop by calculating $\phi = \frac{M(H = 5T)}{M_r}$. While the oxygen stoichiometric sample had $\phi$ = 1.7, our oxygen rich sample yields $\phi$ = 4.0. This higher value of $\phi$ for the oxygen rich sample indicates an increased lack of magnetic saturation in the sample, or in other words, a weakening of long range order.

Another important difference with the oxygen stoichiometric sample is in the virgin curve of the M(H) loop. While the virgin curve of the $\delta = 0$ sample lies entirely within the main loop, our oxygen rich sample shows an unusual magnetic behaviour where a major portion of the virgin curve lies outside the hysteresis loop and meets the main loop only at very high fields.

*A. C. magnetic susceptibility study*

The temperature dependence of the a.c. susceptibility of $CaBaFe_4O_{7.14}$ in the temperature range 20 K – 280 K and at 4 measuring frequencies ranging from 10 Hz to 10 kHz is shown in Fig. 4. The sample shows several interesting features which we will now proceed to discuss separately:

(a) At T = 272 K, one can see a sharp peak which is frequency independent i.e. the position of the peak maximum does not shift with a change in the measuring frequency. This peak corresponds to the paramagnetic to ferrimagnetic (PM-FM) transition occurring in the sample as it is cooled below 272 K. We note here that the oxygen stoichiometric sample also showed a similar peak at around the same temperature. However, in the latter sample, the peak corresponding to the PM-FM transition was the strongest (maximum amplitude) compared to the other peaks. In our sample, this peak is much smaller in magnitude which corresponds to a significant weakening of the magnetic ordering (or a smaller volume fraction of ferrimagnetic domains) which we had mentioned earlier in connection with the M(H) loop.

(b) The oxide $CaBaFe_4O_{7.14}$ shows a broader peak at lower temperature (~ 166 K) which shows pronounced frequency dependence. The peak temperature shifts from 166 K (for a measuring frequency of 10 Hz) to 176 K (for a measuring frequency of 10 kHz). This corresponds to a peak shift of 0.02 per decade of frequency shift ( $p = \frac{\frac{\Delta T_f}{T_f}}{\Delta \log f}$ = 0.02). This value of the parameter $p$ lies within the range for canonical spin glasses, which indicates that this peak is a signature of the sample undergoing a spin glass transition. We confirm this by analyzing the frequency dependence of this peak using the power law form $\tau = \tau_0 \left( \frac{T_f - T_{SG}}{T_{SG}} \right)^{-z\nu}$, where, $\tau_0$ is the shortest relaxation time available to the system, $T_{SG}$ is the underlying spin-glass transition temperature determined by the interactions in the system, $z$ is the dynamic critical exponent and $\nu$ is the critical exponent of the correlation length. The actual fittings were done using the equivalent form of the power law: $\ln \tau = \ln \tau_0 - z\nu \ln \left( \frac{T_f - T_{SG}}{T_{SG}} \right)$. The fit parameters ($\tau_0$ = 7.3 × $10^{-10}$ sec, $z\nu$ = 5.01 and $T_{SG}$ = 162.2 K) give a good linear fit (as can be seen in the inset of Fig. 4), and confirms that this peak does correspond to a spin glass transition in the sample. Whether the magnetic order disappears in the spin glass phase is not clear at the moment. Our data shows that the spin glass transition occurs within the ferrimagnetically ordered phase. Whether this transition is accompanied by the destruction of ferrimagnetic long range order is an open issue as of now. The possibility of coexistence of ferrimagnetic and spin glass orders cannot, however, be ruled out.

(c) At a lower temperature of ~ 110 K a very broad peak is seen (which broadens out to almost kind of a shoulder at lower measuring frequency). We will come back to the nature of this feature in the following section.

*Magnetic field dependence of the spin glass freezing temperature*

In Fig. 5, a.c. susceptibility at 10 kHz driving frequency is plotted as a function of temperature for different external magnetic fields $H_{dc}$ ranging from 0 to 0.3 T. As can be seen from the figure, χ' is suppressed by the magnetic field. First, we focus our discussion on the evolution of the spin glass freezing temperature (marked by a black arrow in the figure), which occurs at ~ 176 K at the lowest applied field. This peak temperature shows a continual shift towards lower temperature as the external magnetic field is increased, and reaches a value of 155 K at an external magnetic field of 0.3 T. Concomitantly, the peak amplitude keeps decreasing as the external magnetic field is increased from 0 to 0.3 T. A further increase of the external magnetic field should eventually suppress the spin glass transition completely.

The purpose of exploring how the freezing temperature responds to external magnetic field is to check the stability of the spin glass system. This is done by examining the field versus temperature phase diagram obtained from the a.c. susceptibility measurement as shown in Fig. 6. As was mentioned above, the spin glass freezing temperature is suppressed by increasing the external magnetic field.

From a theoretical perspective, de Almeida and Thouless [20] studied the Ising spin glass system, and predicted that the spin freezing temperature ($T_g$) depends on $H$. In the low $H$ range, $T_g$ follows the so-called de Almeida-Thouless (AT) line, expressed as $H = H_0\left[1-\frac{T_g(H)}{T_g(0)}\right]^{3/2}$. In addition, Gabay and Toulouse [21] investigated the $H$ dependence of the spin freezing temperature for the Heisenberg spin glass system. This led to the so-called Gabay-Toulouse (GT) line, expressed as $H = H_0\left[1-\frac{T_g(H)}{T_g(0)}\right]^{1/2}$. The AT line and the GT line are the two critical lines predicted in the presence of field on the H-T plane, which mark the phase transition. The first one occurs for an anisotropic Ising spin glass while the second is valid for an isotropic Heisenberg spin glass.

Our sample shows a very interesting behaviour. At low field values ($H_{dc} < 0.15$ T), $T_g$ follows the AT line. This can be seen in Fig. 6, where the red line denotes the AT line. Since the AT line predicts that $T_g \propto H^{2/3}$, so we have plotted $H^{2/3}$ in the H-T phase diagram. However, with an increase in the field (H > 0.15 T), we find deviation from the AT line. Remarkably, it is found that at high field, the variation of $T_g(H)$ agrees with the GT line. This can be seen in the inset of Fig. 6, where the blue line denotes the GT line. Since the GT line predicts that $T_g \propto H^2$, so the H-T phase diagram in the inset is plotted with $H^2$ in the y-axis.

These experimental results can be explained using the theoretical calculation by Kotliar and Sompolinsky [22], who have predicted that in the presence of random anisotropy, the critical behaviour for a spin glass in fields lower than the anisotropy field is close to Ising type following the AT line, and crosses over to Heisenberg behaviour in high fields. The fact that we see a crossover in critical lines on the H-T plane for our sample indicates the existence of magnetic anisotropy in the system. At higher applied fields, the system behaves like a Heisenberg spin glass, where the spins can freeze along any direction with respect to the applied magnetic field. However, when the applied field is lower than the anisotropy field, the spins are forced to be aligned along the local anisotropy axis. The preference of the spin alignment adds an Ising character to the associated spin cluster.

*Domain wall pinning at lower temperature*

In this section, we discuss the third feature seen in the $\chi'$(T) curve – the broad peak at ~ 110 K (Fig. 4). This peak at 110 K does not shift (i.e. the peak maximum occurs at the same temperature) with a change in the external magnetic field $H_{dc}$ (see the red line in Fig. 5). Based on this behaviour, we attribute the origin of the feature seen at ~ 110 K to enhanced domain wall pinning. The signature of this domain wall pinning can also be seen in Fig. 7, where we plot the variation of the coercivity ($H_C$) with temperature. As is clear from the figure, the coercivity is majorly enhanced below 110 K, which occurs due to the domain wall pinning. A close look at the high temperature region, which is enlarged and shown in the inset of Fig. 7, reveals that the coercivity also shows an enhancement below the paramagnetic to ferrimagnetic phase transition temperature (shown by a black arrow), and another enhancement below the spin glass freezing temperature (shown by a blue arrow), as expected.

At this stage, we need to go back to our earlier observation of an unusual initial magnetization curve in the M(H) loop measured at low temperature (Fig. 3). Such unusual magnetic hysteresis behaviour, with the virgin curve lying outside the main hysteresis loop, was earlier associated with irreversible domain wall motion in spinel oxides [23]. Thus, this unusual magnetization curve is an additional confirmation of the domain wall pinning at ~ 110 K that we had mentioned earlier. In fact, we find that the virgin curve lies outside the main M(H) loop for temperatures below 110 K, but above 110 K, it lies completely inside the main hysteresis loop. This is shown in Fig. 8, where we plot the M(H) loops at temperatures slightly below (Fig. 8 (a)), and slightly above (Fig. 8 (b)) 110 K. In the figures, the virgin curves are shown in red for the sake of clarity.

*Origin of the competition between ferrimagnetism and spin glass behaviour*

In order to understand the different magnetic behaviour of $CaBaFe_4O_{7.14}$ with respect to $CaBaFe_4O_7$, we must keep in mind that the oxygen excess in the former induces an increase of the $Fe^{3+}$ content in the structure i.e., the $Fe^{3+}$:$Fe^{2+}$ ratio increases from 1 in the stoichiometric phase to 1.32 in the oxygen hyperstoichiometric phase. As a consequence, the $Fe^{3+}$-$Fe^{3+}$ antiferromagnetic interactions increase in the oxygen rich phase, and may decrease the ferrimagnetism in the structure. Bearing in mind the model previously proposed by Chapon et. al. [4] to explain the competition between 1D ferromagnetism and 2D magnetic frustration in the cobaltite $YBaCo_4O_7$ which has the same hexagonal structure, we must consider the iron framework of our compound. The latter consists of corner-sharing $[Fe_5]_\infty$ bipyramids running along "*c*" interconnected through "$Fe_3$" triangles (Fig. 9). In other words, in both oxides, $CaBaFe_4O_7$ and $CaBaFe_4O_{7.14}$, we can expect, similarly to the hexagonal cobalt oxides $LnBaCo_4O_7$, that the system exhibits an unidimensional magnetic order in the bipyramidal rows along "*c*", whereas the triangular geometry of the iron lattice in the (001) plane induces magnetic frustration as soon as the iron species are coupled antiferromagnetically. Such a model can account for the competition between 1D ferrimagnetism and 2D magnetic frustration in both oxides, $CaBaFe_4O_7$ and $CaBaFe_4O_{7.14}$, and explain that the magnetic frustration may be larger in the latter owing to the appearance of larger short range antiferromagnetic interactions in the (001) plane.

Nevertheless, the valency effect alone is not sufficient to explain the appearance of the spin glass behaviour. Two hypotheses can be considered to explain this particular behaviour. The first scenario deals with the fact that $CaBaFe_4O_{7.14}$ contains interstitial oxygen in spite of the apparent close packed character of the structure, leading to a local puckering of the "$O_4$" and "$BaO_3$" layers. As a result, the distribution of iron in the cationic sites would be locally disordered, leading to a spin glass behaviour. This local distortion would also change the crystal field and would be responsible for the domain wall pinning. The second scenario deals with the fact that the "barium oxygen" framework remains close packed, but that the compound exhibits a cationic deficiency according to the formula $Ca_{0.98}(Ba_{0.98}O_{0.02})Fe_{3.93}O_7$. Such an effect would be similar to that observed for "oxidized" spinels $\gamma$-$Fe_2O_3$ and $Co_{3-x}O_4$, which do not contain interstitial oxygen, but were found to be iron or cobalt deficient [24, 25]. This second scenario would explain the magnetic behaviour of this phase, which is close to that observed for $CaBaFe_{4-x}Li_xO_7$ [26]. In both the systems, the doping of the Fe sites with lithium or vacancies respectively introduces disordering on the Fe sites, which is in turn, at the origin of the appearance of spin glass behaviour at lower temperature. Thus, the competition between 1D ferrimagnetism and spin glass behaviour appears normal. Subsequently, the competition between anisotropic (Ising) and isotropic (Heisenberg) spin glass can be understood from the peculiar geometry of the $[Fe_4]_\infty$ lattice. Finally, the iron vacancies would change the nature of the crystal field in the structure, playing the role of pinning centres. This explains both, the broad peak at 110 K and the enhanced coercivity below this temperature, which are the signatures of domain wall pinning.

The small deviation from the stoichiometry does not allow to distinguish the possibility of interstitial oxygen *vis à vis* that of cationic deficiency from a structural study. Attempts are being made to synthesize similar hexagonal ferrites with larger oxygen excess in order to answer this question.

## Conclusions

This study illustrates the extraordinarily rich physics of the “114” $CaBaFe_4O_{7+\delta}$ ferrite, in connection with its ability to accommodate oxygen excess, similar to what is observed for the spinel family, $Fe_3O_4$ – $\gamma$-$Fe_2O_3$. The remarkable feature of this “114” oxide deals with the competition between ferrimagnetism and spin glass behaviour that can be induced by varying the oxygen content, without changing the hexagonal symmetry of the structure. Such a behaviour can be explained, like for the “114” cobaltites, as due to the competition between 1D magnetic ordering along “*c*” and 2D magnetic frustration in the triangular (001) lattice. Nevertheless, $CaBaFe_4O_7$ differs significantly from $CaBaCo_4O_7$, the latter’s ferrimagnetism originating mainly from a lifting of its 2D geometrical frustration through a strong orthorhombic distortion of its initial hexagonal lattice. We believe that the scenario of cation disordering on iron sites is the key for understanding the magnetism of these materials. Further investigations, especially using neutron diffraction and X-ray synchrotron have to be performed in order to further understand this phenomenon.

## Acknowledgements

The authors acknowledge the CNRS and the Conseil Regional of Basse Normandie for financial support in the frame of Emergence Program. V. P. acknowledges support by the ANR-09-JCJC-0017-01 (Ref: JC09_442369).

**Figure Captions**

**Fig. 1** X-ray diffraction pattern along with the fit for $CaBaFe_4O_{7+\delta}$.

**Fig. 2** $M_{ZFC}$ (T) and $M_{FC}$ (T) curves of $CaBaFe_4O_{7+\delta}$ measured at H = 0.3 T.

**Fig. 3** M(H) curve of $CaBaFe_4O_{7+\delta}$ measured at T = 5 K. The virgin curve is shown in red circles, while the rest of the hysteresis loop is shown in black triangles. The inset shows the M(H) curve of the oxygen stoichiometric sample ($CaBaFe_4O_7$) measured at T = 5 K.

**Fig. 4** The real (in-phase) component of a.c. susceptibility for $CaBaFe_4O_{7+\delta}$ as a function of temperature in the frequency range $f$ = 10 Hz – 10 kHz, at zero static magnetic field ($H_{dc}$) and at a driving ac field ($H_{ac}$) of 10 Oe. The inset shows the plot of ln $\tau$ vs $\ln\left(\frac{T_f - T_{SG}}{T_{SG}}\right)$ for the peak at 166 K.

**Fig. 5** The real (in-phase) component of a.c. susceptibility for $CaBaFe_4O_{7+\delta}$ as a function of temperature. The driving frequency was fixed at $f$ = 10 kHz and $h_{ac}$ = 10 Oe. Each curve was obtained under different applied static magnetic field ($H_{dc}$) ranging from 0 T to 0.3 T.

**Fig. 6** Field vs temperature phase diagram of $CaBaFe_4O_{7+\delta}$. In order to show the AT line, we have plotted $H^{2/3}$ vs $T_g$. The inset shows $H^2$ vs $T_g$ and the GT line.

**Fig. 7** Temperature dependence of coercive field for $CaBaFe_4O_{7+\delta}$. The inset is an enlarged version of the high temperature region.

**Fig. 8** M(H) loops of $CaBaFe_4O_{7+\delta}$ at (a) T = 75 K and (b) T = 135 K.

**Fig. 9** Schematic representation of the $[Fe_4]_\infty$ tetrahedral framework of hexagonal $CaBaFe_4O_{7+\delta}$ showing the $Fe_5$ bipyramids sharing corners with $Fe_3$ triangular groups (adapted from Ref. 11).

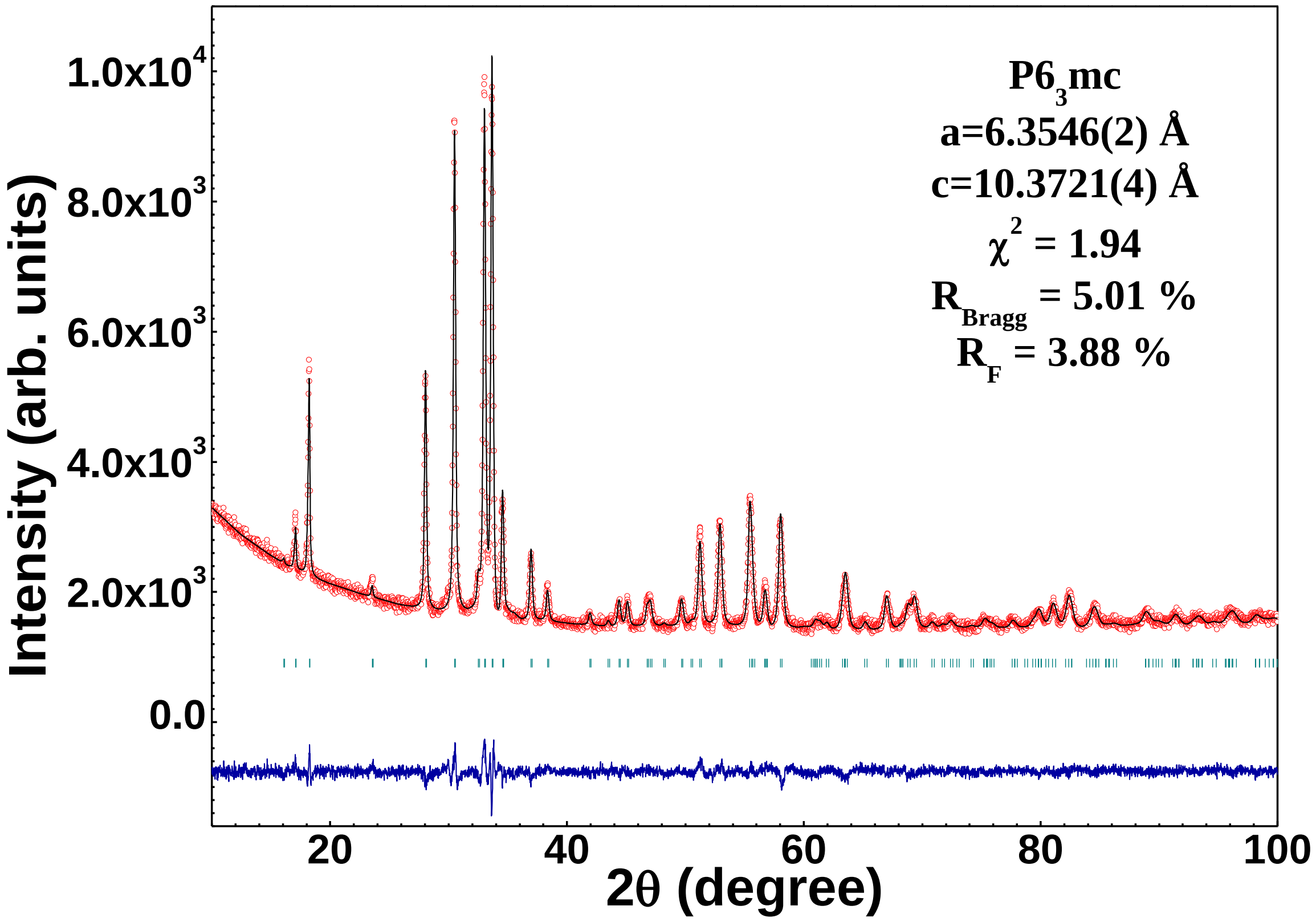

**Fig. 1**. X-ray diffraction pattern along with the fit for $CaBaFe_4O_{7+\delta}$.

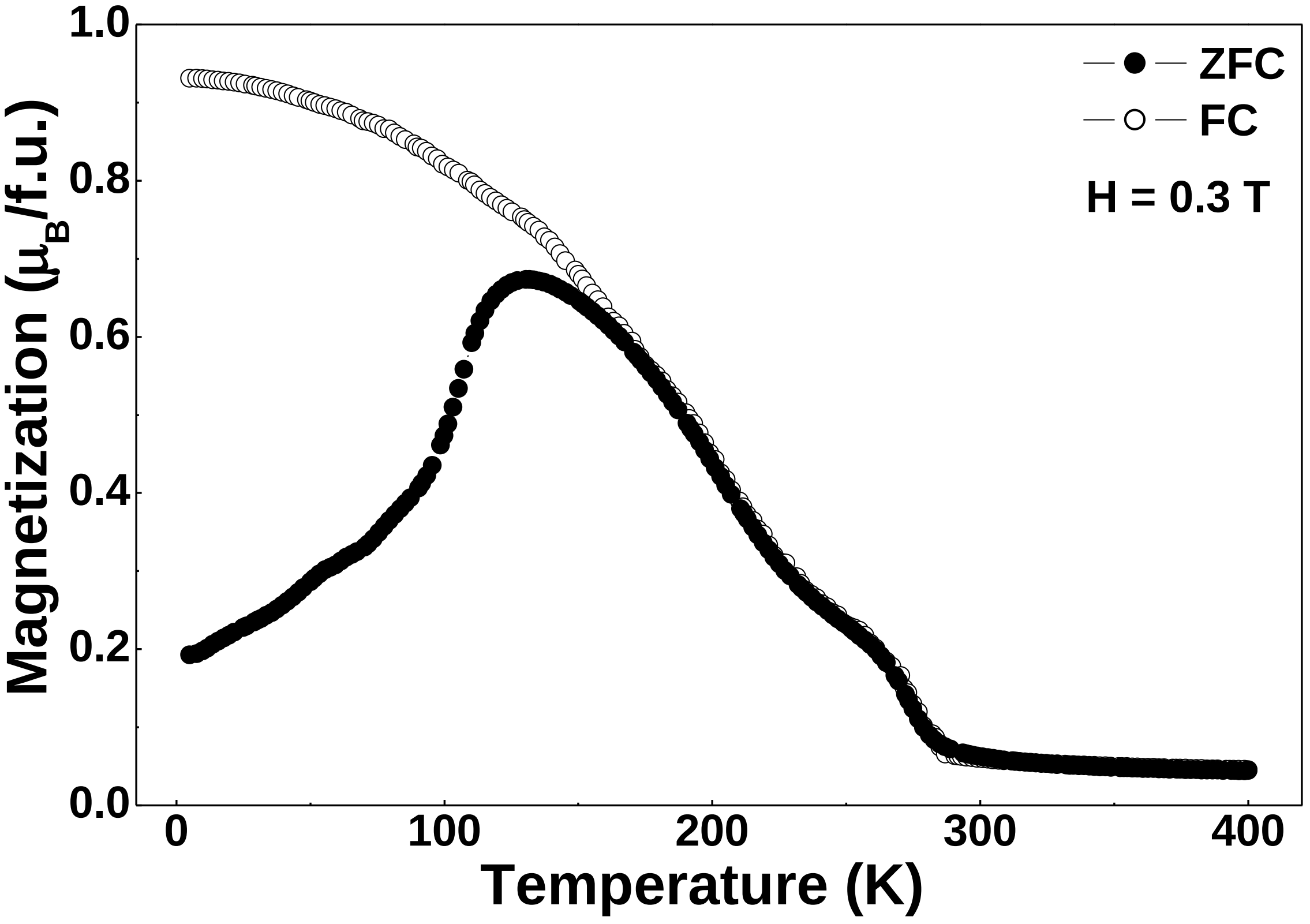

**Fig. 2**. $M_{ZFC}$ (T) and $M_{FC}$ (T) curves of $CaBaFe_4O_{7+\delta}$ measured at H = 0.3 T.

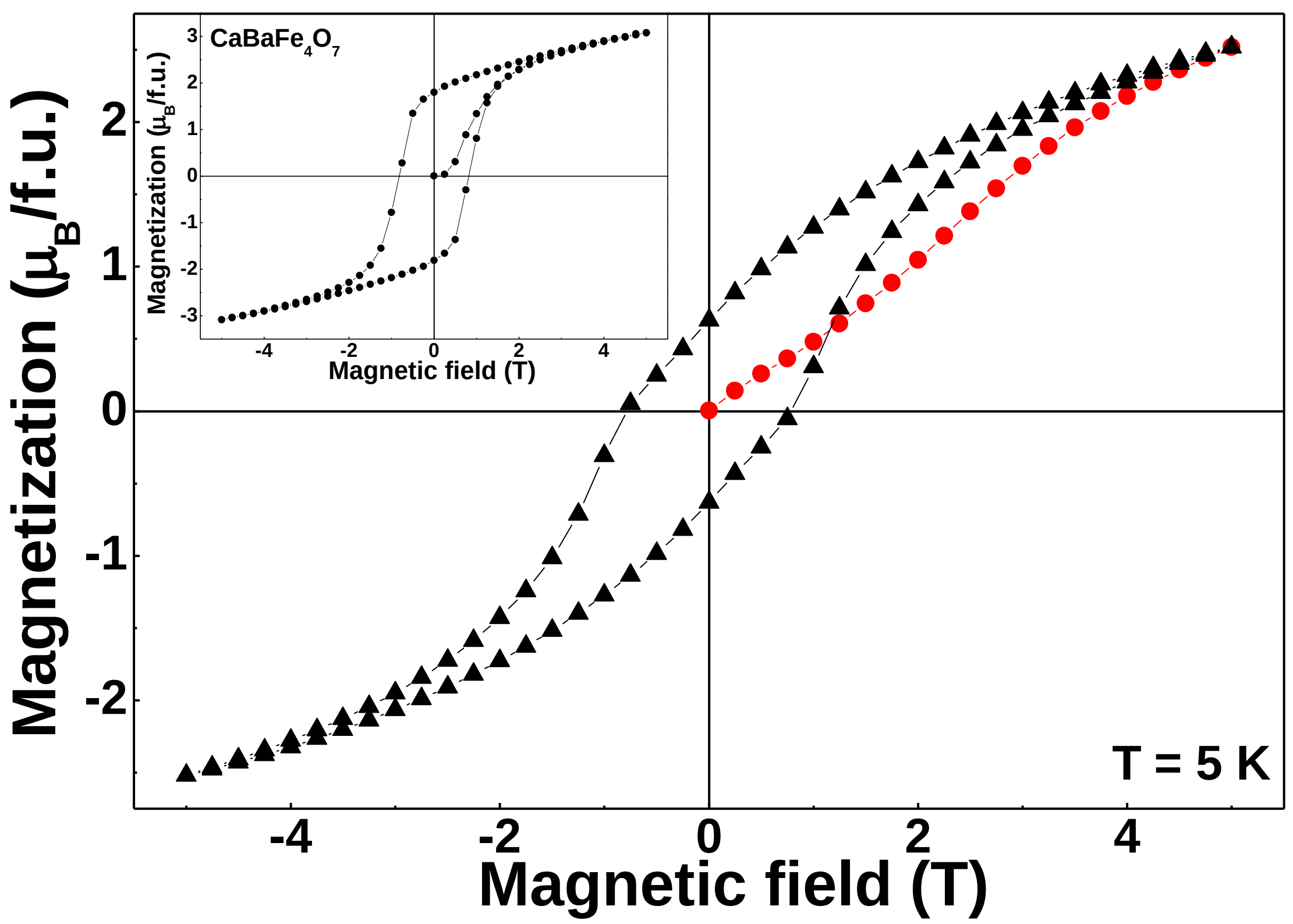

**Fig. 3.** M(H) curve of $CaBaFe_4O_{7+\delta}$ measured at T = 5 K. The virgin curve is shown in red circles, while the rest of the hysteresis loop is shown in black triangles. The inset shows the M(H) curve of the oxygen stoichiometric sample ($CaBaFe_4O_7$) measured at T = 5 K.

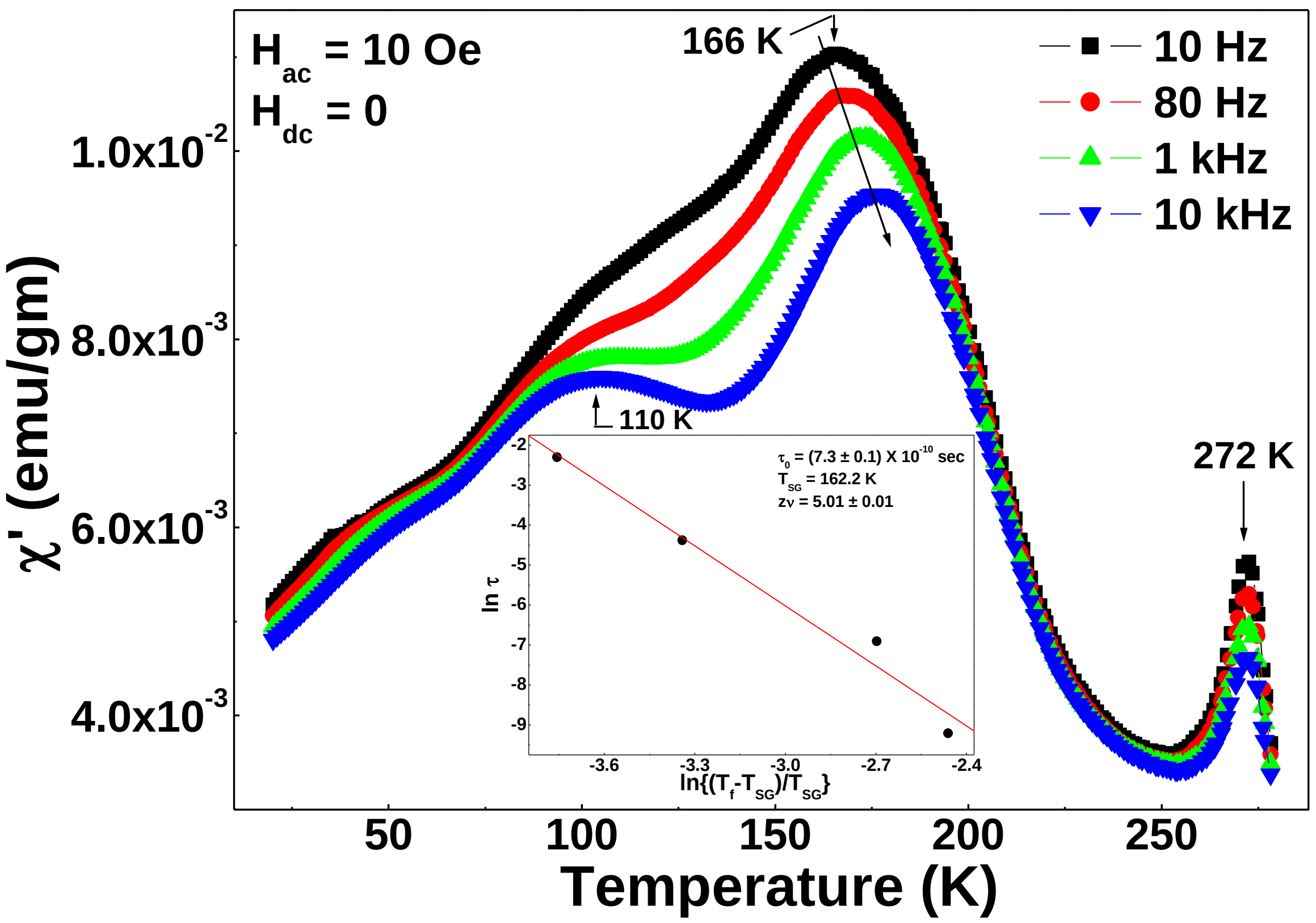

**Fig. 4**. The real (in-phase) component of a.c. susceptibility for $CaBaFe_4O_{7+\delta}$ as a function of temperature in the frequency range $f$ = 10 Hz – 10 kHz, at zero static magnetic field ($H_{dc}$) and at a driving ac field ($H_{ac}$) of 10 Oe. The inset shows the plot of ln $\tau$ vs $\ln\left(\frac{T_f - T_{SG}}{T_{SG}}\right)$ for the peak at 166 K.

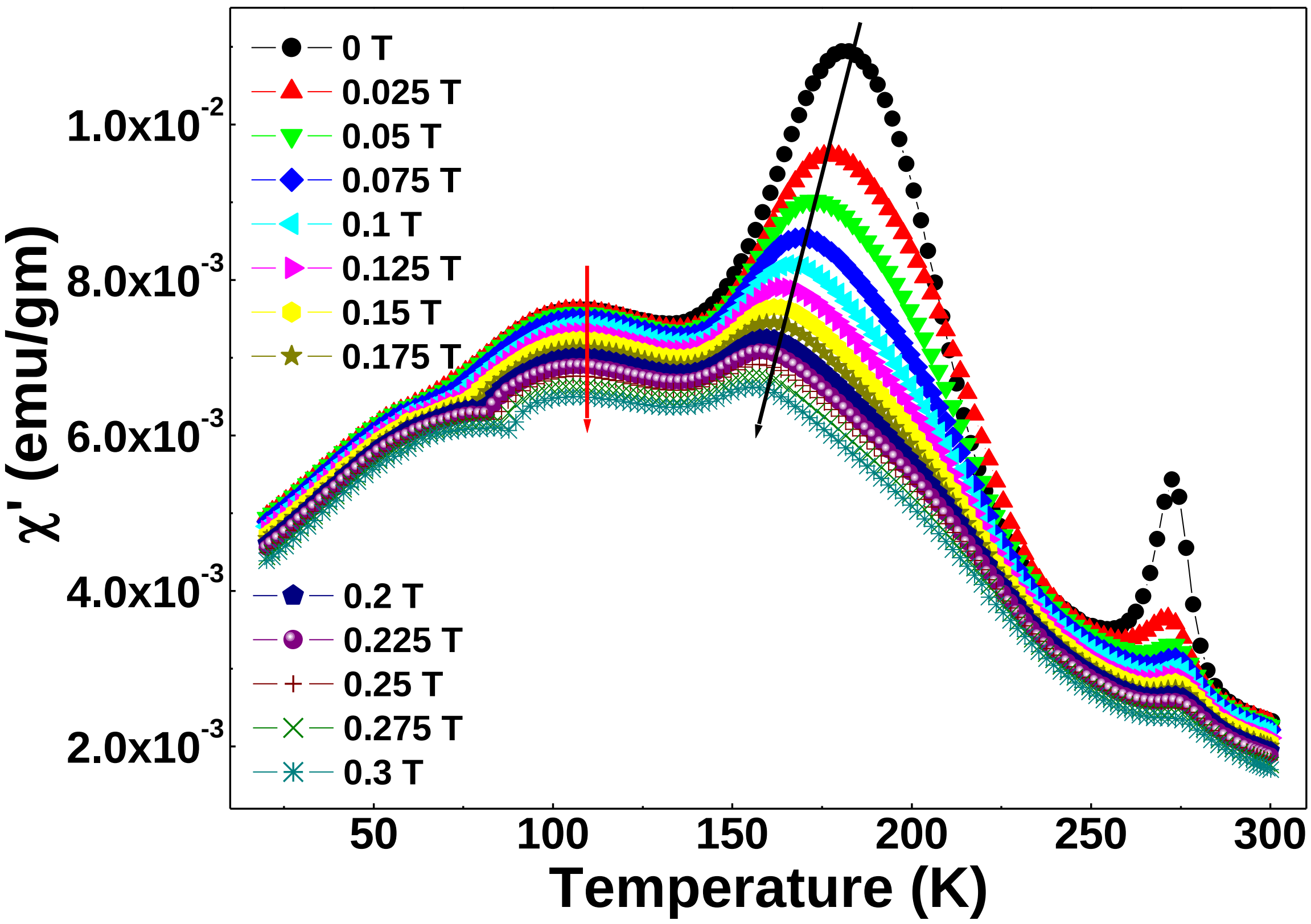

**Fig. 5**. The real (in-phase) component of a.c. susceptibility for $CaBaFe_4O_{7+\delta}$ as a function of temperature. The driving frequency was fixed at $f$ = 10 kHz and $h_{ac}$ = 10 Oe. Each curve was obtained under different applied static magnetic field ($H_{dc}$) ranging from 0 T to 0.3 T.

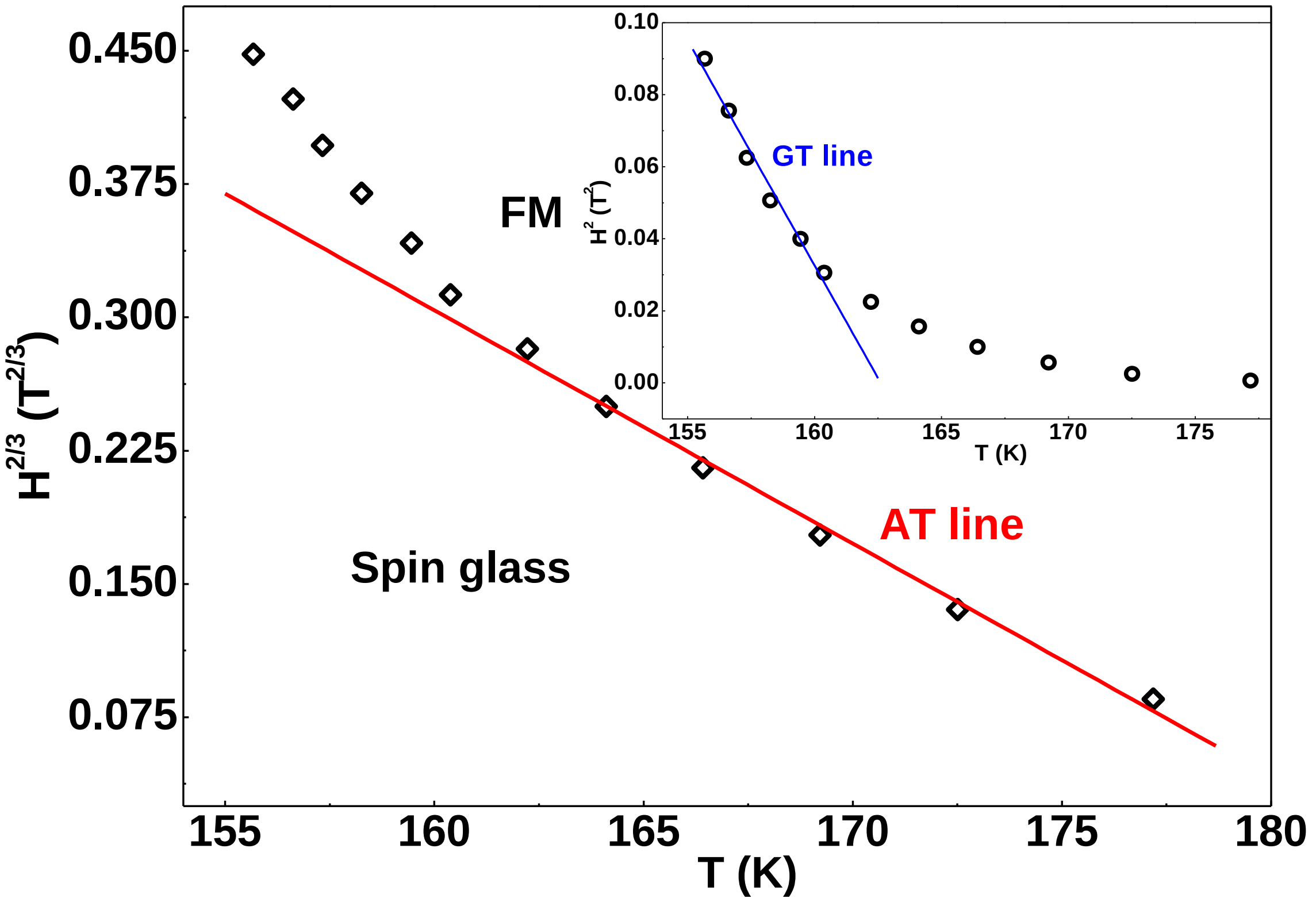

**Fig. 6**. Field vs temperature phase diagram of $CaBaFe_4O_{7+\delta}$. In order to show the AT line, we have plotted $H^{2/3}$ vs $T_g$. The inset shows $H^2$ vs $T_g$ and the GT line.

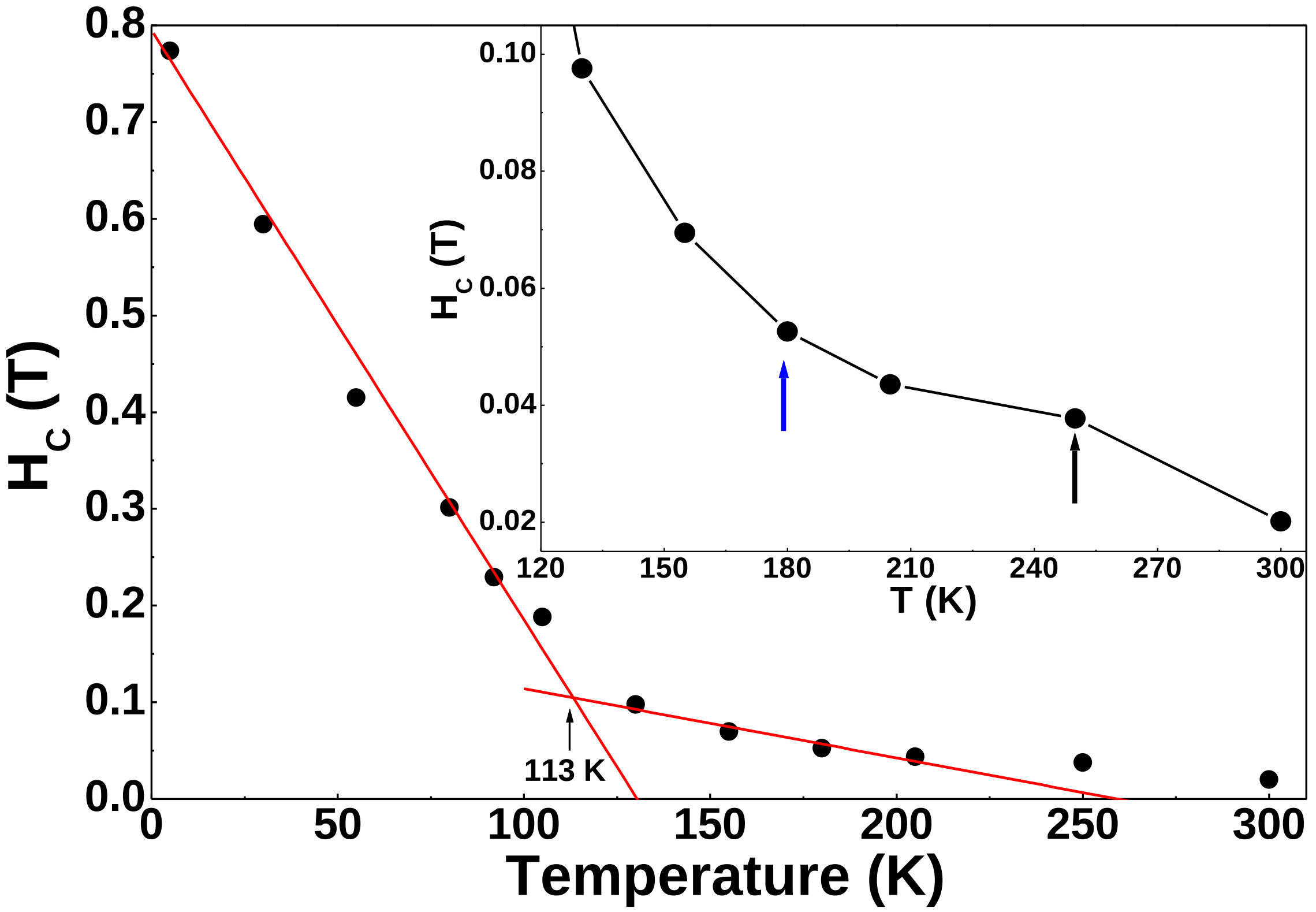

**Fig. 7**. Temperature dependence of coercive field for $CaBaFe_4O_{7+\delta}$. The inset is an enlarged version of the high temperature region.

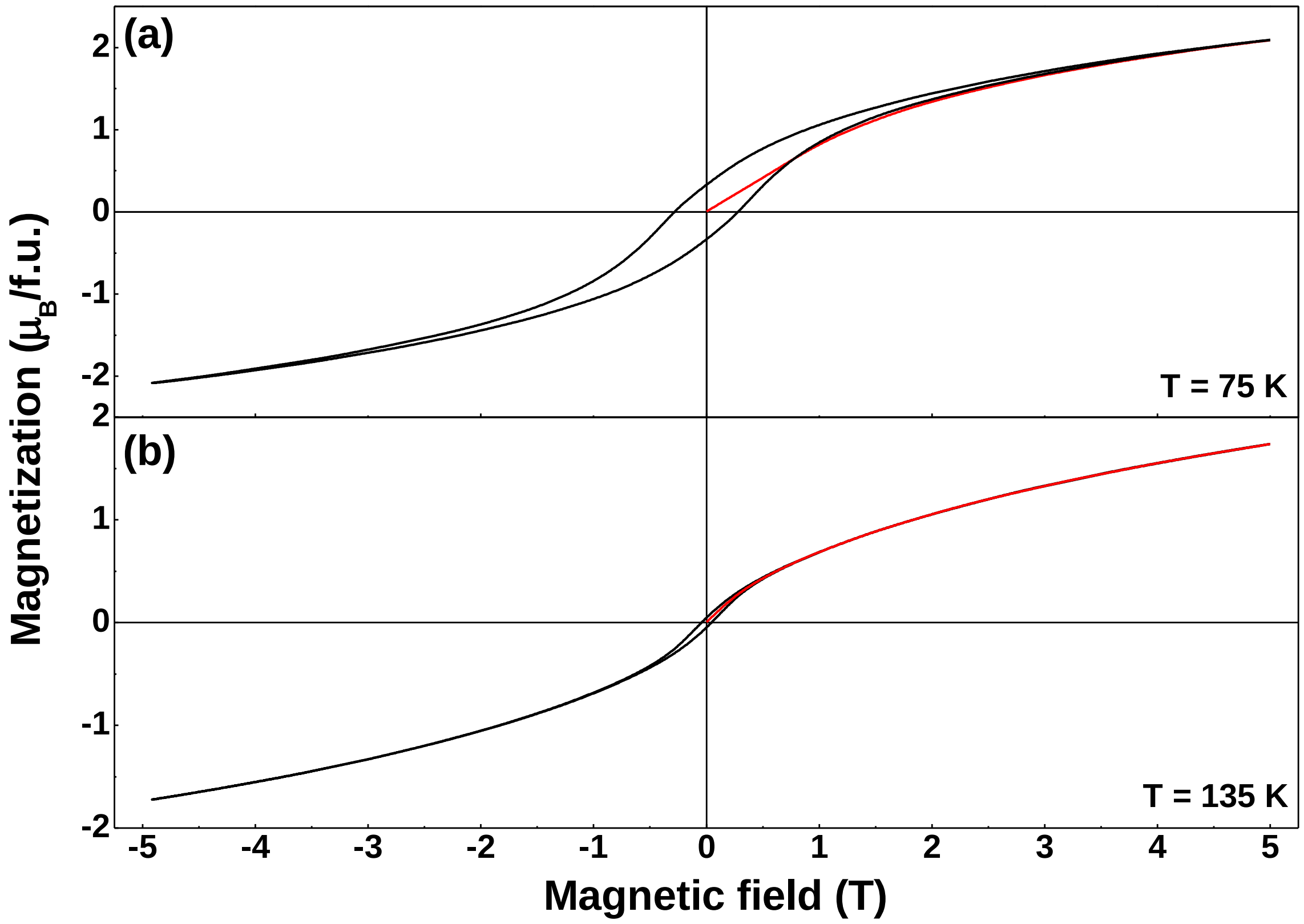

**Fig. 8**. M(H) loops of $CaBaFe_4O_{7+\delta}$ at (a) T = 75 K and (b) T = 135 K.

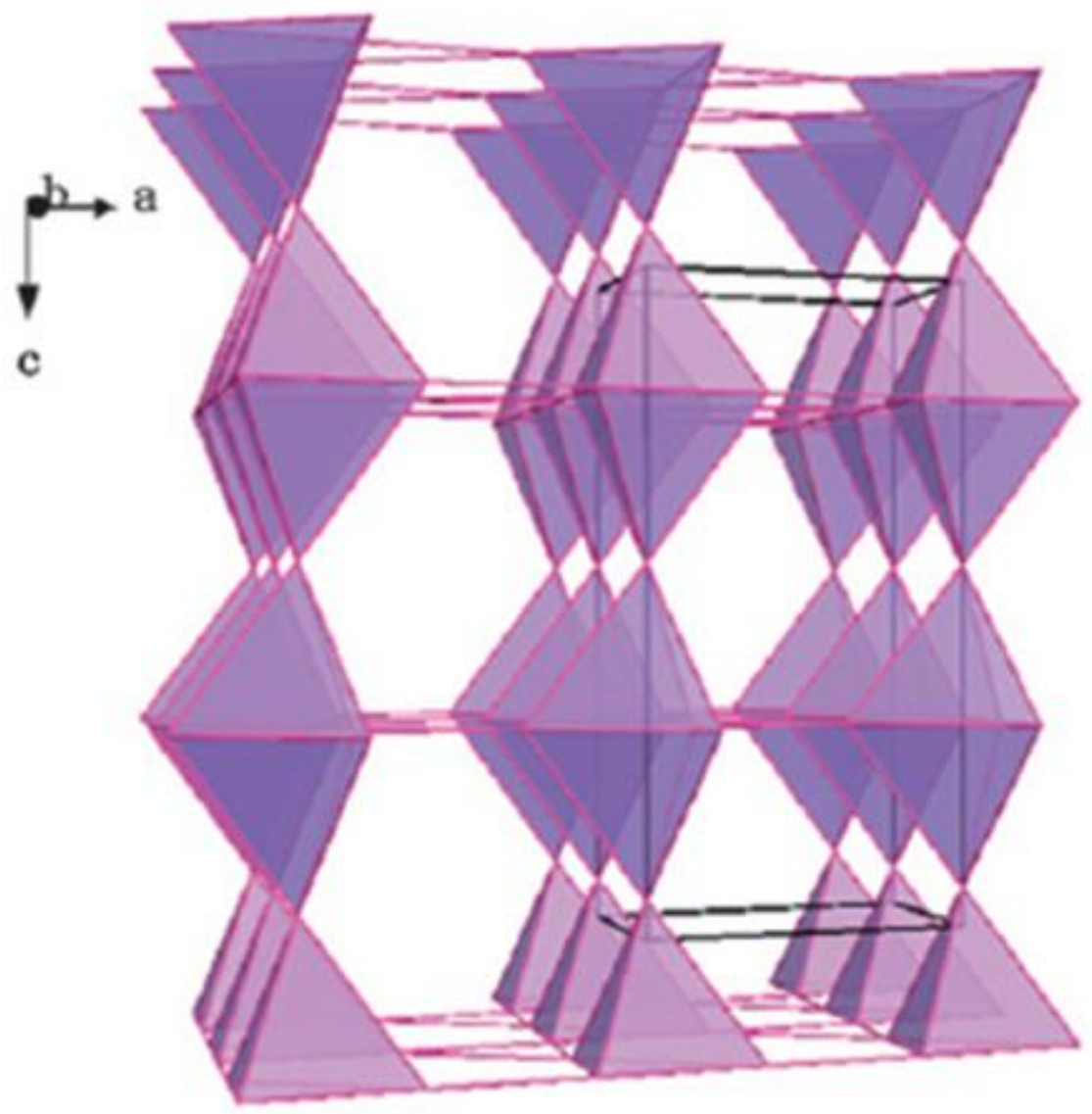

**Fig. 9**. Schematic representation of the $[Fe_4]_\infty$ tetrahedral framework of hexagonal $CaBaFe_4O_{7+\delta}$ showing the $Fe_5$ bipyramids sharing corners with $Fe_3$ triangular groups (adapted from Ref. 11).